\documentclass{cpbtex}
\usepackage{array}
\usepackage{float}
\usepackage{booktabs}
\usepackage{textcomp}
\usepackage{multirow}
\usepackage{amstext}
\usepackage{graphicx}
\usepackage{color}
\usepackage{ulem}
\usepackage{hyperref}

\def\NNO{NiNb$_2$O$_6$}
\def\CNO{CoNb$_2$O$_6$}

\begin{document}
\begin{CJK*}{GBK}{song}

\title{Semiclassical approach to spin dynamics of a ferromagnetic $S=1$ chain\thanks{Project supported by the National Key R\&D Program of China (Grant No. 2023YFA1406500), and the National Science Foundation of China (Grant Nos. 12334008 and 12174441)}}


\author{Chengchen Li
	$^{1}$,
    Yi Cui
    $^{1,2}$,
    Weiqiang Yu
    $^{1,2}$,
    \ and \ Rong Yu
	$^{1,2}$\thanks{Corresponding author. E-mail:rong.yu@ruc.edu.cn}\\
$^{1}$Department of Physics and Beijing Key Laboratory of Opto-electronic Functional Materials\\
 and Micro-nano Devices, Renmin University of
China, Beijing 100872, China\\
$^{2}$Key Laboratory of Quantum State Construction and Manipulation (Ministry of Education),\\ Renmin University of China, Beijing, 100872, China}  

\date{\today}
\maketitle

\begin{abstract}
Motivated by recent experimental progress in the quasi-one-dimensional quantum magnet \NNO, we study the spin dynamics of an $S=1$ ferromagnetic Heisenberg chain with single-ion anisotropy by using a semiclassical molecular dynamics approach. This system undergoes a quantum phase transition from a ferromagnetic to a paramagnetic state under a transverse magnetic field, and the magnetic responses reflecting this transition is well described by our semiclassical method. We show that at low-temperature the transverse component of the dynamical structure factor depicts clearly the magnon dispersion, and the longitudinal component exhibits two continua associated with single- and two-magnon excitations, respectively. These spin excitation spectra show interesting temperature dependence as effects of magnon interactions. Our findings shed light on experimental detection of spin excitations in a large class of quasi-one-dimensional magnets.
\end{abstract}



\section{Introduction}\label{Sec:Intro}

Quantum fluctuations in quasi-one-dimensional (q1D) quantum spin systems can give rise to a number of
novel quantum phases and exotic emergent phenomena, including fractionalized excitations, symmetry protected topological order, and emergent quantum integrability~\cite{Giamarchibook_2004, Chen_Science_2012, Cui_PRL_2019, Faure_NP_2018, Wang_Nature_2018, Zou_PRL_2021, Nikitin_NC_2021, MotrunichNUM2019, Xi_CPB_2022}. Many interesting phenomena are associated with quantum antiferromagnetic (AFM) systems for the strong quantum fluctuations in these systems: On the one hand, the AFM order parameter does not commute with the Heisenberg or XXZ Hamiltonian. On the other hand, frustration effects are more significant in AFM systems. As for ferromagnetic (FM) systems, many properties have been well studied. For example, the ground state of a FM Heisenberg model is a simple product state with all spins aligned in parallel, and the elementary excitation is a spin wave, or a magnon, which carries a spin quantum number $S=1$.\cite{QuantumMagnetismBook}

Recently, there are numerous experimental and theoretical studies on a q1D FM quantum magnet \CNO.\cite{Coldea_Science_2010, Lee_NP_2010, Kinross_PRX_2014, Morris_NP_2021, Fava_PNAS_2020, Chernyshev_arXiv_2312.03829} This compound consists of effective $S=1/2$ FM chains that are aligned in a frustrated manner (forming a triangular lattice in the plane perpendicular to the chain direction). Along each chain, the system is approximately described by an Ising model, and the elementary excitations are domain walls, or spinons. This is further evidenced by the experimental finding that, upon applying a transverse field, the low-energy properties of the system is controlled by a 1D quantum critical point (QCP) belonging to the transverse field Ising universality, which is hidden inside a 3D magnetically ordered phase~\cite{Kinross_PRX_2014}.

\CNO\ is one compound in a large family of q1D magnets~\cite{Kristallogr_1976}. Recently, its $S=1$ counterpart, \NNO, has been studied~\cite{Chauhan_PRL_2020, Xu_PRX_2022}. Similar to \CNO, \NNO\ also exhibits Ising anisotropy, and the ground state undergoes a FM to paramagnetic (PM) quantum phase transition under a transverse field~\cite{Heid_JPCM_1996, Yaeger_PR_1977}. Thermal transport measurements found similar behaviors of the thermal conductivity near the quantum phase transitions of both \CNO\ and \NNO.\cite{Xu_PRX_2022} However, recent THz spectroscopy measurements~\cite{Chauhan_PRL_2020} revealed a magnon-like excitation mode in \NNO, implying very different underlying physics of these two materials. Theoretical study showed that the effective model of \NNO\ is the FM $S=1$ Heisenberg chain, and the Ising anisotropy comes from the single-ion anisotropy, which is weak compared to the Heisenberg exchange interaction~\cite{Chauhan_PRL_2020, Sharma_PRB_2022, Nayak_PRB_2022}. As a result, the elementary excitation of the \NNO\ system is a magnon. It has been demonstrated based on experiment and spin-wave theory, that magnon-magnon interaction plays a crucial role in understanding the temperature dependence of the spin gap in \NNO.\cite{Sharma_PRB_2022} One then wonder the spectral signatures of single and multiple magnons, and how the spin excitation spectrum would evolve with field and temperature. The linear spin-wave theory works well at zero temperature and in the zero and large field limits. However, it would be complicated to apply this theory to finite temperature, and the applicability of the theory to generic field values has to be justified.

Motivated by recent progress on \NNO\ and the open theoretical questions, in this work, we examine the spin dynamics of an effective 1D spin model for \NNO\ via a simple semiclassical approach, which combines the numerical techniques of classical Monte Carlo (MC) and molecular dynamics (MD) simulations. This method allows us to access to spin dynamics of the system on both sides of the quantum phase transition. We calculate the spin dynamical structure factor (DSF) and confirm a magnon dispersion in the transverse component of the DSF. Remarkably, we find two continua excitations in the longitudinal channel at finite temperatures, and show that they are associated with single- and two-magnon excitations, respectively. Our study reveal the importance of the single-ion anisotropy for the the observed two-magnon continuum. We discuss ways in probing these spectral signatures in \NNO\ and other q1D magnets with large spin size.

\section{Model and Methods}\label{Sec:Model}

We study an effective 1D $S=1$ FM Heisenberg model for \NNO. The model Hamiltonian reads as
\begin{equation} \label{Eq:Ham}
H=-J\sum_{i}\mathbf{S}_{i}\cdot \mathbf{S}_{i+1} -D\sum_{i}(S^z_{i})^2-gh_x\sum_{i}S^x_{i},
\end{equation}
where $J>0$ is the FM interaction along the chain, $D>0$ is the easy-axis single-ion anisotropy, and $h_x$ denotes the applied transverse magnetic field with gyromagnetic factor $g$. For \NNO, $J\sim 10$-$15$ K, $D\sim5$, and $g\approx2.3$.\cite{Heid_JPCM_1996, Yaeger_PR_1977} In this work, we set $J=1$ as the energy unit and take $D=0.5$ in the numerical simulations.

We investigate the spin dynamics of the model in Eq.~\eqref{Eq:Ham} via a semiclassical method by using classical MC and MD simulations~\cite{Landau_PRB_1994,Landau_PRB_1990}. An individual simulation consists of two major stages as described below.

{First, we treat the spin as a classical vector with fixed magnitude $S=1$ and introduce two angles $\theta$ and $\phi$ to express the direction of the spin vector in the three-dimensional spherical coordinates as $\mathbf{S}=(\sin\theta\cos\phi,\sin\theta\sin\phi,\cos\theta)$. Then a classical MC simulation} is employed to obtain the thermodynamic properties of the system and to generate spin configurations in equilibrium for the MD simulation. We then sample in the $(\cos\theta ,\phi)$ space to ensure the uniform sampling in the entire spin configuration space. The variation $(\delta\cos\theta ,\delta\phi)$ is adjusted during the thermalization of a simulation so that the acceptance rate stays close to $50\%$ within the temperature range we considered. We further implement an overrelaxation algorithm~\cite{Brown_PRL_1987,Creutz_PRD_1987} to reduce the autocorrelation time. During every overrelaxation, we perform a specular reflection of the local spin $\mathbf{S}_i$ about the local field $\mathbf{h}_i$ exerted on it within the plane spanned by $S_i^z$ and $\mathbf{h}_i$. In practice, an overrelaxation is performed when the spin configuration is not accepted at a MC update.

During the MC simulation, we calculate the magnetic susceptibility $\chi$ with respect to the applied field via
\begin{align}\label{Eq:chi}
\chi&=\frac{1}{NT}(<(\sum_{i}S^x_i)^2>-<\sum_{i}S^x_i>^2)
\end{align}
where $T$ is the temperature and $N$ represents the number of total spins.

In the next step, we obtain the spin dynamics of the system via solving the equation of motion (EOM) for each spin:
\begin{equation}\label{Eq:EOM}
\frac{d}{dt}\mathbf{S}_i(t)=i[H(t),\mathbf{S}_i(t)],
\end{equation}
where $\mathbf{S}_i(t)=e^{iHt} \mathbf{S}_i e^{-iHt}$ is the spin operator in the Heisenberg representation. We solve the above EOM via the MD simulation. To be specific, we start from the spin configuration generated by the MC algorithm. $\mathbf{S}_i(t)$ of each spin is then determined from the integration of coupled EOMs in Eq.~\eqref{Eq:EOM}.\cite{Watson_PR_1969} These equations are numerically integrated by a vectorized fourth-order predictor-corrector method~\cite{Numerical_Analysis}. We carry the integration up to a maximum time $t_{max}=100/J$, with a time step $\Delta=0.01/J$.

The DSF $S(\mathbf{q},\omega)$ is then calculated via
\begin{align}\label{Eq:DSF}
S^{\alpha\alpha} (\mathbf{q},\omega) &= \sum_{i,j} e^{i\mathbf{q} \cdot (\mathbf{r}_i-\mathbf{r}_j)} \int_{-\infty}^{\infty} \left[\langle S^\alpha_i(t) S^\alpha_j(0)\rangle -\langle S^\alpha_i(t)\rangle \langle S^\alpha_j(0)\rangle \right] e^{i\omega t} dt,
\end{align}
where $\alpha=x,y,z$. Note that we have used Simpson's rule~\cite{Numerical_Analysis} for the numerical integration involved in Eq.~\eqref{Eq:DSF}, and for the time integration in Eq.~\eqref{Eq:DSF}, we took the time interval to be 0.1$J^{-1}$. To obtain good statistics, each DSF is averaged over $100$ MD simulations with independent spin configurations. In the simulation, we have taken a chain of length $L=100$ and with periodic boundary condition. The finite-size effect is found to be negligible.

We compare our numerical results with those of the linear spin-wave theory, which we outline below.
We first treat the $h_x=0$ case in the FM state. From the Hamiltonian in Eq.~\eqref{Eq:Ham}, we perform a Holstein-Primakoff (H-P) transformation to map the spin Hamiltonian to that of interacting bosons (magnons).
We then get the magnon dispersion
\begin{equation}\label{Eq:dispz}
 \epsilon_q = 2J(1-\cos q) + D.
\end{equation}

In the PM state driven by the transverse field $h_x$, we can construct a spin-wave theory in the large $h_x$ limit. In this limit, the ground state corresponds to a configuration that all spins are fully polarized to the $S^x$ direction by the transverse field. The elementary excitation is then a magnon (spin wave) where one spin deviates from the ground-state configuration.
To see this explicitly, we first perform a rotation of $\pi/2$ about the $S^y_i$ axis for each spin. This sends $S^x_i\rightarrow S^z_i$ and $S^z_i\rightarrow -S^x_i$. We then make use of
\begin{equation}\label{Eq:Sx2}
 (S^x_i)^2 = \frac{1}{4}[(S^+_i)^2 + (S^-_i)^2] + \frac{1}{2} [S^2-(S^z_i)^2].
\end{equation}
The Hamiltonian in the rotated basis is then rewritten as
\begin{equation}\label{Eq:HamRot}
 H = -J\sum_{i}\mathbf{S}_{i}\cdot \mathbf{S}_{i+1} -gh_x\sum_{i}S^z_{i} +\frac{D}{2}\sum_{i}(S^z_{i})^2 -\frac{D}{4} \sum_i [(S^+_i)^2+(S^-_i)^2],
\end{equation}
We can then perform a H-P transformation in this rotated basis. The transverse field $h_x$ (longitudinal field in the rotated basis) corresponds to the chemical potential of the magnons. The penultimate term acts as a repulsive Coulomb interaction between magnons, and the last term refers to creating and annihilating two magnons on a site. In the single-magnon sector, the spin-wave (single-magnon) dispersion reads
\begin{equation}\label{Eq:dispx}
 \epsilon_q = 2J(1-\cos q) - D/2 +gh_x.
\end{equation}
The energy-momentum relation of the two-magnon continuum reads
\begin{equation}\label{Eq:TwoMagnon}
 \omega_2(q,k) = \epsilon_{k+q}+\epsilon_{k} = 4J[1-\cos \frac{q}{2} \cos (k+\frac{q}{2})] - D +2gh_x.
\end{equation}

\section{Phase diagram and field evolution of the spin gap}\label{Sec:PhD}
\begin{figure}[thbp]
\centering
\includegraphics[width=12.0cm]{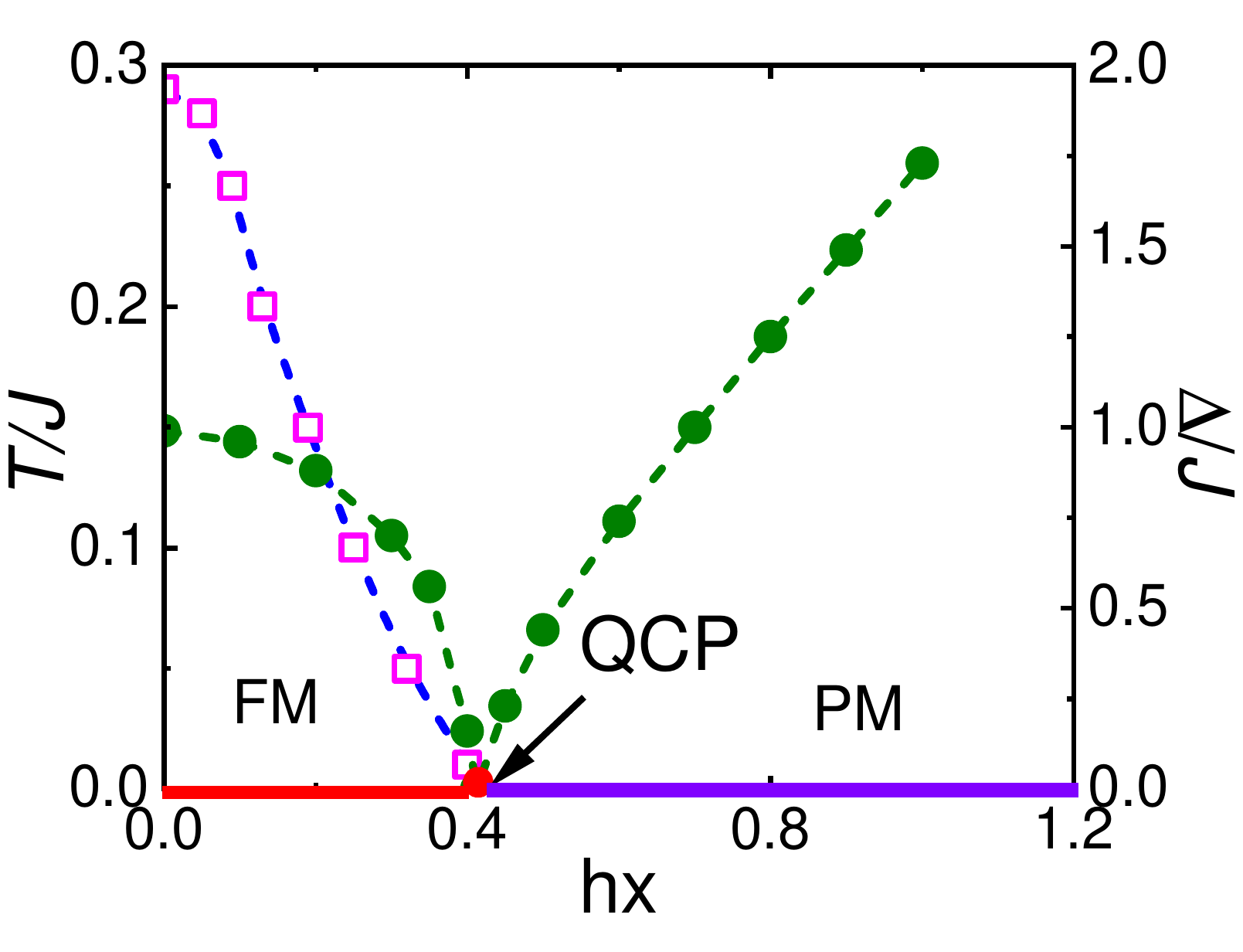}
\caption{\label{fig:phase} The phase diagram in the field-temperature plane and the field dependent spin gap. The green curve with circles represents the spin gap extracted from the transverse DSF $S^{yy}(q,\omega)$ at $q=0$ and $T=0.01$. The dashed line with magenta squares shows a crossover between classical renormalized and disordered regimes, which is determined from the peak position of the field dependent susceptibility (see Fig.~\ref{fig:hx}). These field dependent behaviors indicate a FM to PM quantum phase transition with QCP located at $h_x^c\approx0.4$.}
\end{figure}

\begin{figure}[thbp]
\centering
\includegraphics[width=15.0cm]{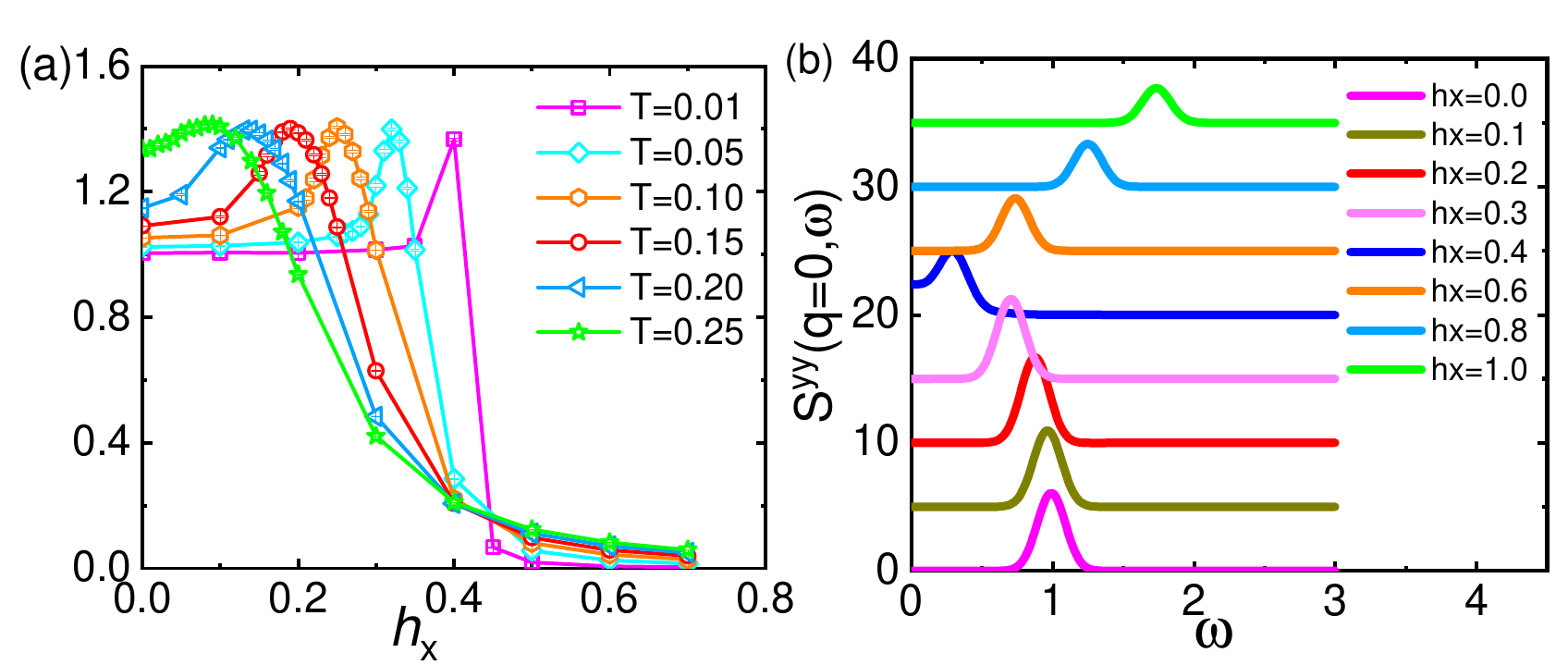}
\caption{\label{fig:hx} (a): Field dependent magnetic susceptibility $\chi$ at several temperatures. The peak indicates a crossover between the classical renormalized and disordered regimes. (b): The transverse DSF $S^{yy}(q,\omega)$ at $q=0$ and $T=0.01$. The energy of the peak in each curve determines the spin gap shown in Fig.~\ref{fig:phase}.
}
\end{figure}

At zero field, the model in Eq.~\eqref{Eq:Ham} has a FM Ising ordered ground state. Because of the single-ion anisotropy $D$, the spins are lined up along the $S^z$ direction. The transverse field $h_x$ suppresses the Ising order by aligning spins in the $S^x$ direction, and turns the FM ground state to a PM one via a quantum phase transition at $h_x^c$ (see Fig.~\ref{fig:phase}). In a 1D system, the FM order is stabilized only at $T=0$, but finite-temperature crossovers can still be observed. {For example, at low temperature on the FM side, the system behaves as a classically FM ordered one because of the largely enhanced correlation length. On the other hand, on the PM side, the transverse field introduces fluctuations to substantially suppress the FM correlation. We then expect a finite-temperature crossover between a renormalized classical and a disordered regime when increasing the field. To see this, we} calculated the field dependent transverse susceptibility $\chi$ from the MC simulations at several temperatures. As shown in Fig.~\ref{fig:hx}(a), with increasing $h_x$, $\chi$ first increases, develops a peak, then is quickly suppressed. The peak signals a crossover from the classical renormalized regime to the disordered regime, shown as the dashed line with square symbols in Fig.~\ref{fig:phase}. Note that quantum fluctuations are not captured by our classical MC simulations, therefore the quantum critical regime in the vicinity of a QCP is not identified. However, we can determine the position of the QCP by extrapolating the crossover line to the $T=0$ limit. This gives $h_x^c/J\approx0.4$.

We then calculate the transverse component of the DSF $S^{yy}(q=0,\omega)$ at $T=0.01$ via the semiclassical MD method. As shown in Fig.~\ref{fig:hx}(b), $S^{yy}(q=0,\omega)$ shows a single peak at finite energy $\omega$, indicating a gapped excitation. At this low temperature, the spins are converged to their ground-state configuration. It is then expected that the DSF is characterized by the excitation from the ground state to the lowest excited state. The extracted spin gap (in Fig.~\ref{fig:phase}) first decreases with increasing the field $h_x$ until it closes at $h_x/J\approx0.4$. The gap then reopens in higher fields. The gap closing and reopening behavior is consistent with that of a FM transverse field Ising chain, demonstrating clearly the FM to PM transition in the ground state. The critical field value extrapolated from the DSF agrees well with that from the susceptibility. {Given that the elementary excitations on both FM and PM sides are magnons as shown in Sec.~\ref{Sec:Model}~~, the gap closing and reopening behavior clearly indicates the softening of magnons when the FM-to-PM transition is appraoched.}

\section{Spin excitation spectra}\label{Sec:Sqw}
\begin{figure}[thbp]
\centering
\includegraphics[width=15.0cm]{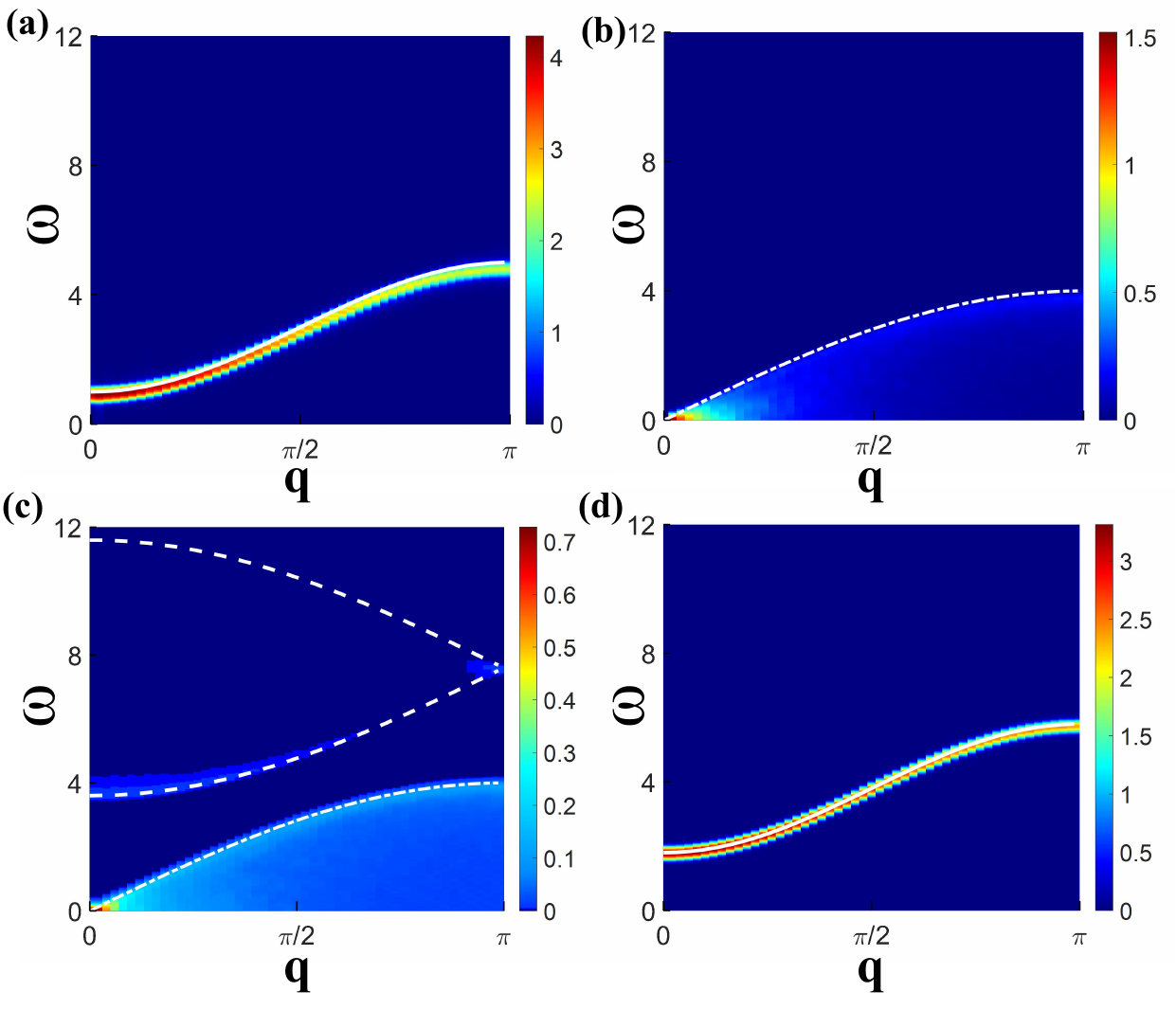}
\caption{\label{fig:sqw}
Spin excitation spectra of $S^{xx}(q,\omega)$ [in (a) and (c)] and $S^{zz}(q,\omega)$ [in (b) and (d)] at $T=0.1$ for $h_x=0$ (upper panels) and $h_x=1$ (lower panels). The solid lines depict single magnon (spin wave) dispersion. The dash-dotted lines refer to the boundaries of the continua associated with one-magnon excitations. The dashed lines show boundaries of the two-magnon continua. See text for more information.
}
\end{figure}

With the phase diagram settled, we now examine the low-energy excitations of the system. At zero temperature, the low-energy spin excitations in either the FM or the PM phase is described by the spin-wave theory presented in Sec.~\ref{Sec:Model}~~~. The elementary excitation is a magnon which characterizes the deviation from the fully polarized spin state along either $S_z$ or $S_x$ direction, depending on the field. In Fig.~\ref{fig:sqw}, we present calculated $S^{xx}(q,\omega)$ and $S^{zz}(q,\omega)$ at finite momentum $q$ for two field values $h_x=0$ and $h_x=1$ at $T=0.1$. We see that the transverse mode, $S^{xx}(q,\omega)$ in the FM and $S^{zz}(q,\omega)$ in the PM state, displays clearly a dispersive magnon band.
However, by fitting the numerical magnon dispersions to Eqs.~\eqref{Eq:dispz} and \eqref{Eq:dispx}, we find they better fit to $\epsilon_q = 2J(1-\cos q) + 2D$ at $h_x=0$ and
$\epsilon_q = 2J(1-\cos q) - D +gh_x$ at $h_x=1$. Compared to the spin-wave theory, $D$ is doubled.
The discrepancy comes from the semiclassical approximation to the single-ion anisotropy term. The EOM for $S^x_i$ associated with this term reads,
$\frac{d}{dt}S^{x}_i(t)=i[D(S^z_i)^2,S^{x}_i] = D(S^y_i S^z_i + S^z_i S^y_i)$. For $S=1$, the term on the right-hand side corresponds to a quadrupolar operator. However, by treating the spin operators as classical vectors, we take the following approximation $D(S^y_i S^z_i + S^z_i S^y_i)\approx 2D S^y_i S^z_i$ when performing the MD simulation. This approximation causes the double counting of the $D$ term, as appeared in the magnon dispersion in MD simulations. Note that this issue only leads to a global shift of the spin gap, but will not affect the shape of the dispersion.

Next, we investigate the longitudinal modes, which correspond to $S^{zz}(q,\omega)$ in the FM and $S^{xx}(q,\omega)$ in the PM state. These modes exhibit more interesting features. As shown in Fig.~\ref{fig:sqw}(b) and (c), for both $h_x=0$ and $h_x=1$, the longitudinal modes exhibit prominent gapless continua extending to the Brillouin zone (BZ) boundary. This is rather surprising. First, a continuum is usually associated to fractionalized spin excitations, but for a FM Heisenberg chain, the elementary excitation is a magnon carrying $S=1$. Moreover, even the continuum of an Ising chain is gapped when the system is away from the QCP.

To understand the nature of these continuous excitations, we note that the spectral weights of the longitudinal excitations are much weaker than those of the transverse modes. This implies that the observed continua are finite-temperature effects. At finite temperature, the magnon density is nonzero. These magnons reside in the single magnon band, and can be further excited. Assuming a magnon residing on the state with momentum $k$ and energy $\epsilon_k$, it can be excited to another state in the magnon band with momentum $k+q$ and energy $\epsilon_{k+q}$. The excitation energy is then
\begin{equation}\label{Eq:Continuum}
 \omega(q,k) = \epsilon_{k+q}-\epsilon_{k} = 4J\sin \frac{q}{2} \sin (k+\frac{q}{2}),
\end{equation}
forming a gapless continuum with half-bandwidth of $4J$. Note that this process only alters the energy and momentum of a magnon, but does not modify the magnon density. So it shows up in the longitudinal channel. Also note that it relies on a finite magnon density in the initial state, which is possible only at finite temperature. This type of excitations is exactly the gapless continua observed in Fig.~\ref{fig:sqw}(b) and (c), on both the FM and PM sides of the phase diagram.

Besides this gapless continuum, we observe another gapped continuum in the longitudinal mode at $h_x=1$ on the PM side. Numerically, we find it fits to the energy-momentum relation of two-magnon excitations in Eq.~\eqref{Eq:TwoMagnon} when we substitute $2D$ for $D$. This implies that this continuum originates from two-magnon excitations. But we need to understand why these excitations are only observed in the PM phase and in the longitudinal channel.

The answer to these questions lies in the Hamiltonian in the rotated basis of Eq.~\eqref{Eq:HamRot}. The last term of this Hamiltonian contains a $(S^+_i)^2$ process, which generates a two-magnon bound state (with a pair of magnons on one site) from the fully polarized (zero magnon) state. As a result, the ground state is a superposition of the zero- and two-magnon states. It is then possible that the magnon density does not change when exciting from the ground state to the two-magnon continuum, which must take place in the longitudinal channel. This is indeed what is observed in the numerical results. In the FM state, from Eq.~\eqref{Eq:Ham} we see that the $(S^+_i)^2$ process is missing in the Hamiltonian. The two-magnon state is not mixed to the ground state and hence a two-magnon continuum is not observed.

\begin{figure}[thbp]
\centering
\includegraphics[width=15.0cm]{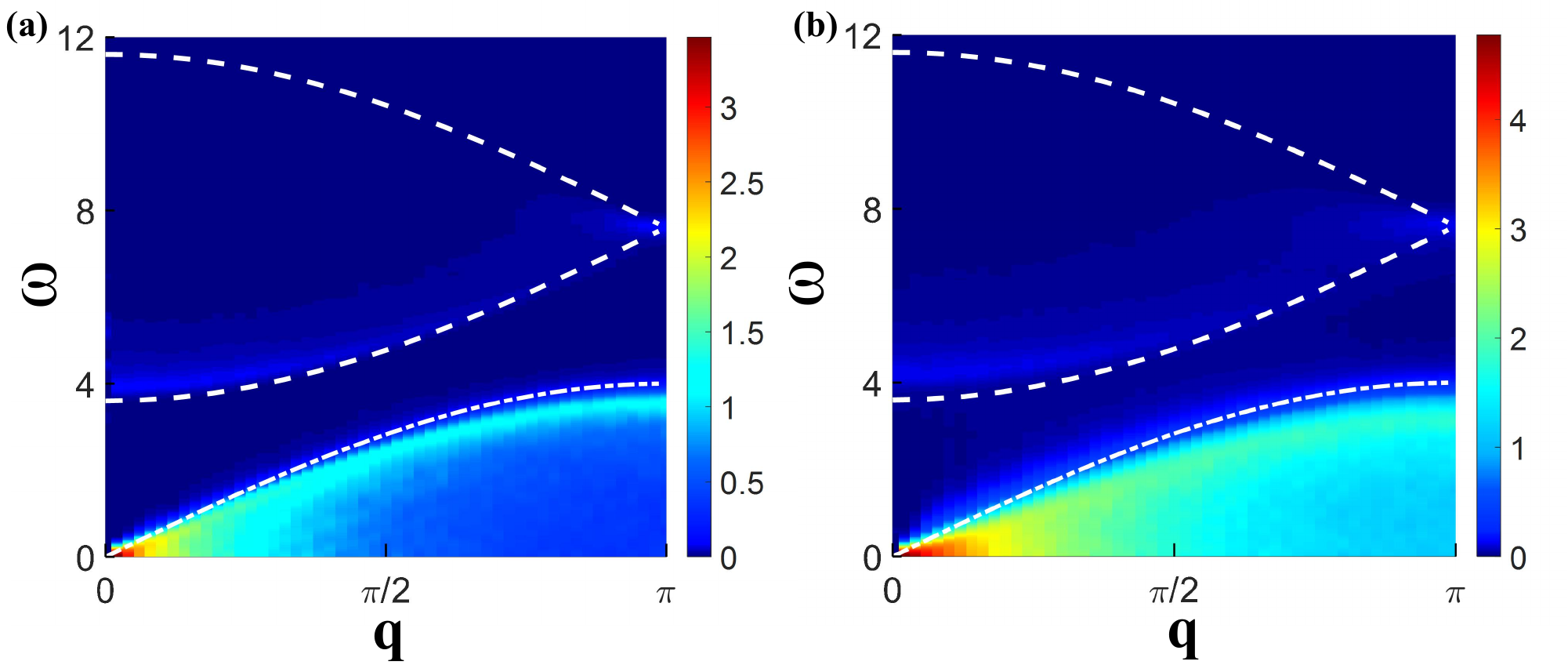}
\caption{\label{fig:sqw2} $S^{xx}(q,\omega)$ at $h_x=1$ and temperatures $T=0.5$ [in (a)] and $T=1.0$ [in (b)], respectively. The dashed and dash-dotted lines show boundaries of magnon continua at zero temperature according to the spin-wave theory.}
\end{figure}

Next we examine the evolution of these continua in the longitudinal excitations with temperature. In Fig.~\ref{fig:sqw2} we show $S^{xx}(q,\omega)$ at $h_x=1$ in the PM state for two temperatures $T=0.5$ and $T=1$. Compared to $S^{xx}(q,\omega)$ at $T=0.1$ in Fig.~\ref{fig:sqw}(c), we see two major features: First, with increasing temperature, the overall spectral weights initially increases, then decreases; second, the bandwidths of both continua decrease with increasing temperature. The initial increase of spectral weights of the continua is due to the larger (multi-)magnon population at higher temperature. The spectral weights then decay as temperature futher increases. The second feature is associated with the magnon-magnon interaction. As we discussed in Sec.\ref{Sec:Model}~~~, the $\frac{D}{2}(S^z_i)^2$ term in Eq.~\eqref{Eq:HamRot} acts as a repulsive onsite interaction between magnons. With increasing temperature, the localization effects caused by this term leads to suppression of both bandwidth and spectral weight of the transverse single-magnon mode (not shown). Accordingly, the bandwidths of the two continua in the longitudinal mode are also suppressed. For the two-magnon continuum, since the bandwidth suppression is about the center of the continuum, effectively, the two-magnon spin gap at $q=0$ increases with increasing temperature. {In addition, from Eq.~\eqref{Eq:TwoMagnon} we see that the center of the two-magnon continuum depends on both $D$ and $h_x$. They also affect the evolution of the two-magnon spin gap.}

\section{Discussions and conclusions}\label{Sec:Disc}
In this work, we have investigated spin dynamics of an $S=1$ ferromagnetic Heisenberg model with single-ion anisotropy by using a semiclassical MD method. We show that the magnetic responses and evolution of the spin gap with the applied transverse magnetic field is well described by our numerical approach. We further study the DSF of this model, and find a magnon dispersion in the transverse component. Interestingly, we find two continua excitations in the longitudinal component of DSF at finite temperature. Based on spin-wave analysis, these two continua are ascribed to single- and two-magnon excitations, respectively. We show that the single-ion anisotropy couples the 2-magnon bound state to the 0-magnon ground state, which is crucial for the observation of the two-magnon continuum in the longitudinal excitation spectra. Note that this mechanism does not exist in the $S=1/2$ model, but is generic for Heisenberg models with $S\geqslant 1$. {In the present work we have shown the existence of magnon-origin continua in the case of the easy-axis anisotropy (positive $D$ in our model). It would be interesting to further explore the similar physics in the easy-plane anisotropic case (negative $D$) by using advanced quantum many-body techniques, given that either the Haldane or the large-$D$ ground state in the easy-plane case is beyond our present semiclassical approach.} Our model is pertinent to the $S=1$ spin chain compound \NNO\ and other q1D magnets with large spin size~\cite{Curley_PRM_2021}. Our results also suggest that a very different low-energy spin dynamics in \NNO\ from its effective spin $S=1/2$ counterpart, \CNO, though both compounds exhibit easy-axis anisotropy. The calculated spin dynamics, especially the continua in spin excitation spectra can be verified by various experiments including inelastic neutron scattering and Raman scattering.

\addcontentsline{toc}{chapter}{Acknowledgment}
\section*{Acknowledgment}
We thank J. Dai for useful discussions. Financial supports are given in the footnote on the first page.

\end{CJK*}  
\end{document}